\documentclass[aps, prd, twocolumn, superscriptaddress, nofootinbib, preprintnumbers]{revtex4}
\usepackage{graphicx}
\usepackage{amsmath}
\usepackage{mediabb}                     %   [For PDF graphics] Automatic PDF-bounding box

\allowdisplaybreaks[1]
\usepackage{ulem} %\sout{hogehoge}
 \usepackage{color}

\newcommand{\GeV}{\ensuremath{\,\text{GeV} }}

\begin{document}
\preprint{KUNS-2560}
%%%%%%%%%%%%%%%%%%%% TITLE %%%%%%%%%%%%%%%%%%%%
\title{
Landau pole in the Standard Model with 
%electroweakly charged 
weakly interacting scalar fields
}  
\date{\today}

\author{Yuta Hamada}
\affiliation{
Department of Physics, Kyoto University, Kyoto 606-8502, Japan
}
\author{Kiyoharu Kawana}
\affiliation{
Department of Physics, Kyoto University, Kyoto 606-8502, Japan
}
\author{Koji Tsumura}
\affiliation{
Department of Physics, Kyoto University, Kyoto 606-8502, Japan
}

%%%%%%%%%%%%%%%%%%%% ABSTRACT %%%%%%%%%%%%%%%%%%%%

\begin{abstract}
We consider the Standard Model with a new scalar field $X$ 
which is a $n_X^{}$ representation of the $SU(2)_L$ with a hypercharge $Y_X$. 
The renormalization group running effects on the new scalar quartic coupling constants are evaluated. 
Even if we set the scalar quartic coupling constants to be zero at the scale of the new scalar field, 
the coupling constants are induced by the one-loop effect of the weak gauge bosons. 
Once non-vanishing couplings are generated, the couplings rapidly increase 
by renormalization group effect of the quartic coupling constant itself. 
As a result, the Landau pole appears below Planck scale if $n_X^{}\geq 4$. 
We find that the scale of the obtained Landau pole is much lower than that evaluated  
by solving the one-loop beta function of the gauge coupling constants. 
\end{abstract}

\maketitle

%%%%%%%%%%%%%%%%%%%% INTRODUCTION %%%%%%%%%%%%%%%%%%%%

% Higgs mass and Plank Scale
The discovery of the Higgs boson~\cite{Aad:2012tfa,Chatrchyan:2012ufa} and the determination of its mass 
open up the possibility that the Standard Model (SM) can be valid up to very high energy scale such as 
string/Planck scale~\cite{Holthausen:2011aa,Bezrukov:2012sa,Degrassi:2012ry,Alekhin:2012py,Masina:2012tz,
Hamada:2012bp,Jegerlehner:2013cta,Jegerlehner:2013nna,Hamada:2013cta,Buttazzo:2013uya,Branchina:2013jra,
Kobakhidze:2014xda,Spencer-Smith:2014woa,Branchina:2014rva}.\footnote{
See Refs.~\cite{Foot:2007iy,Meissner:2007xv,Iso:2009ss,Iso:2009nw,Hamada:2014xka,Haba:2014sia,
Kawana:2014zxa,Haba:2014oxa,Kawana:2015tka} for the analysis in simple extensions of the SM. 
}
In particular, the Higgs self-coupling constant and its beta function almost vanish at the same time around 
Planck scale if the top-quark mass $M_t$ is $171\GeV$. 
%, which means that the SM Higgs potential becomes very flat. 
This fact may imply relations between physics at the weak scale and physics at Planck scale, 
e.g.,  Higgs inflation~\cite{Bezrukov:2007ep,Salvio:2013rja,Hamada:2013mya,Hamada:2014iga,Bezrukov:2014bra,
Hamada:2014wna,Hamada:2014raa,Hamada:2015ria},
%\footnote{Non-monotinic $\epsilon$ allow to predict large tensor to scalar ratio~\cite{BenDayan:2009kv}.} 
a multiple point criticality principle~\cite{Froggatt:1995rt} and a maximum entropy principle~\cite{Kawai:2011qb,Kawai:2013wwa,Hamada:2014ofa,Kawana:2014vra,Hamada:2014xra}.

% Higgs and New Physics ( extra Higgs field )
On the other hand, we know that the SM should be extended to include dark matter, the origin of neutrino masses, 
and a mechanism for generating the baryon number of the universe. 
Such an extension often introduces a new scalar field $X$ at around the electroweak/TeV  or 
some intermediate scale which is charged under the $SU(2)_L\times U(1)_Y$ gauge group. 
If we add a sufficiently large $SU(2)_L$ isospin multiplet, the new field provides a good dark matter candidate 
because direct interactions with SM particles are forbidden automatically by renomalizability~\cite{Cirelli:2005uq,Cirelli:2007xd,Cirelli:2009uv}. 
It is known as the Type-II seesaw mechanism that Majorana masses for neutrinos can be generated 
by introducing a vacuum expectation value of a scalar triplet with $Y=1$~\cite{Konetschny:1977bn,Magg:1980ut,
Cheng:1980qt,Schechter:1980gr}. 
It is also known that additional scalar boson loops can increase the triple (SM-like) Higgs boson coupling\cite{Kanemura:2004ch,Kanemura:2004mg}, 
which help to satisfy the sphaleron decoupling condition for the electroweak baryogenesis\cite{Kuzmin:1985mm,
Cohen:1993nk,Quiros:1994dr}. 
A special choice of the Higgs multiplet, the Higgs septet with $Y=2$~\cite{Hisano:2013sn, Kanemura:2013mc}, 
can give sizable deviations in 
the observed Higgs boson couplings without conflicting the stringent constraint from the electroweak $\rho$ parameter. 
There are so many extended models motivated by various motivations, so that it is important to constrain such possibilities 
in a generic way. 

% Landau pole
The triviality bound\cite{Cabibbo:1979ay,Lindner:1985uk} is studied to give such a generic constraint 
on the extended Higgs models\cite{Flores:1982pr,Bovier:1984rh}. 
Considering the renormalization group equation (RGE) for the Higgs quartic coupling constant, 
the energy dependent coupling constant $\lambda(Q^2)$ grows with an energy scale $Q$. 
At the end of the day, the running coupling constant blows up at a certain scale $\Lambda_\text{LP}$. 
The scale is called Landau pole (LP), 
where the coupling constant becomes infinite $1/\lambda(\Lambda_\text{LP})=0$. 
The absence of the LP at a certain scale, e.g., Plank scale, is often required 
to constrain the Higgs quartic coupling 
(equivalently the Higgs boson mass in the SM)\cite{Cabibbo:1979ay,Lindner:1985uk}. 
This bound can be understood as a criteria for perturbativity of the theory. 
%\red{In other words, the theory is valid up to the LP.} 

% In this letter, ...
In this letter, we give the RGE analysis of the scalar quartic coupling constants 
in the SM with one more scalar multiplet $X$, where the field  
$X$ is a $n_X^{}$ representation under $SU(2)_L$ with a hypercharge $Y_X$. 
The model predicts the LP below Planck scale for $n_X\geq 4$ 
if $X$ appears in the electroweak/TeV scale, where
the scale of the LP is defined by the blowup of a coupling constant. 
The LP we will derive from the scalar quartic coupling constants 
is much lower than that obtained by solving one-loop beta functions 
of the gauge coupling constants~\cite{DiLuzio:2015oha}. 
The point is that the quartic coupling constants of $X$ are rapidly induced 
by the electroweak gauge couplings with large coefficients (due to the large electroweak charges) 
in the beta functions, even if the initial values of the quartic coupling constants are set to be zeros. 
Once a finite quartic coupling is injected, the RGE running of the quartic coupling constant leads the LP. 
In general, there are many degrees of freedom to choose the initial condition for the new coupling constants. 
Among them, we evaluate a conservative scale (the largest scale) of the LP.  
~\\

%%%%%%%%%%%%%%%%%%%% MODEL %%%%%%%%%%%%%%%%%%%%

The scalar potential of the model is given by
%----------------------------------------
\begin{align}
V=
&-\mu^2 \Phi^\dag \Phi +\lambda(\Phi^\dag\Phi)^2 +M_X^2X^\dag X +\lambda_X|X^\dag X|^2 \nonumber \\
&+\kappa|X^\dag X||\Phi^\dag \Phi|+\kappa'(X^\dag T^a_X X)(\Phi^\dag T^a_\Phi \Phi) \nonumber \\
&+\lambda_X'(X^\dag T^a_X X)^2+\lambda_X''(X^\dag T^a_XT^b_X X)^2+\cdots,
\end{align}
%----------------------------------------
where $\Phi$ is the SM Higgs doublet, $X$ is a new scalar field, 
and the $SU(2)_L$ generator for $X$ is $T_X^a (a=1,2,3)$. % of the $n_X^{}$ representation. 
Among the scalar quartic coupling constants, some of them are related each other 
depending on the electroweak charges of $X$. 
For quadruplets, extra coupling constants are allowed,
%----------------------------------------
\begin{align}
& \lambda_{\Phi {\Phi^\dag}^2 X} \Phi \Phi^\dag \Phi^\dag X
=\lambda_{\Phi {\Phi^\dag}^2 X} \Phi^i \Phi^\dag_j \Phi^\dag_k X^{i' j k} \epsilon_{i i'}, 
\\
&\lambda_{{\Phi^\dag}^2X^2} {\Phi^\dag}^2X^2
=\lambda_{{\Phi^\dag}^2X^2} \Phi^\dag_i \Phi^\dag_j X^{i k \ell} X^{j k' \ell'} \epsilon_{kk'} \epsilon_{\ell\ell'},\\
&\lambda_{{\Phi^\dag}^3 X} {\Phi^\dag}^3 X
=\lambda_{{\Phi^\dag}^3 X} \Phi^\dag_i \Phi^\dag_j \Phi^\dag_k X^{i j k}, %\epsilon_{i i'} \epsilon_{j j'} \epsilon_{k k'}
\end{align}
%----------------------------------------
where we adopt the symmetric tensor notation, i.e., 
$(X^{111}, X^{112}, X^{122}, X^{222})=(X^1, X^2/\sqrt{3}, X^3/\sqrt{3}, X^4)$ for $n_X=4$. 
The former two coupling constants exist for $Y_X=1/2$, while the last does for $Y_X=3/2$. 
There is a similar coupling of $\lambda_{{\Phi^\dag}^2X^2}$ for a sextet. 
Note that there is an accidental global $U(1)$ symmetry, if all the additional dimension four couplings are forbidden. 
We summarize the independent coupling constants and possible dimension four extra coupling constants 
of each model in Table~\ref{Table:coupling}. 
%
%++++++++++++++++++++++++++++++++++++++++++++++++++
\begin{table}[tb]
%\begin{center}
\begin{tabular}{|c|c|c|}
\hline
$(n_X^{},Y_X)$ & independent couplings & dim-$4$ extra couplings \\ \hline
$(4,1/2)$ & $\lambda, \kappa, \kappa', \lambda_X, \lambda_X'$ & ${\Phi^\dag}^2 X^2,~ \Phi {\Phi^\dag}^2 X$\\ \hline
$(4,3/2) $& $\lambda, \kappa, \kappa', \lambda_X, \lambda_X'$ & ${\Phi^\dag}^3X$\\ \hline
$(5,\text{real})$ & $\lambda, \kappa, \lambda_X$ & $\times$\\ \hline
%$(5,0)$ & $\lambda, \kappa, \kappa', \lambda_X, \lambda_X', \lambda_X''$ & $\Phi\Phi^\dag X^2$\\ \hline
$(5,1)$ & $\lambda, \kappa, \kappa', \lambda_X, \lambda_X', \lambda_X''$ & $\times$\\ \hline
$(5,2)$ & $\lambda, \kappa, \kappa', \lambda_X, \lambda_X', \lambda_X''$ & $\times$\\ \hline
$(6,1/2)$ & $\lambda, \kappa, \kappa', \lambda_X, \lambda_X', \lambda_X''$ & ${\Phi^\dag}^2 X^2$\\ \hline
$(6,3/2)$ & $\lambda, \kappa, \kappa', \lambda_X, \lambda_X', \lambda_X''$ & $\times$\\ \hline
$(6,5/2)$ & $\lambda, \kappa, \kappa', \lambda_X, \lambda_X', \lambda_X''$ & $\times$\\ \hline
$(7,\text{real})$ & $\lambda, \kappa, \lambda_X, \lambda_X''$ & $\times$\\ \hline
%$(7,0)$ & $\lambda, \kappa, \kappa', \lambda_X, \lambda_X',\lambda_X'', \lambda_X'''$& $ \Phi\Phi^\dag X^2$\\ \hline
$(7,1)$ & $\lambda, \kappa, \kappa', \lambda_X, \lambda_X',\lambda_X'', \lambda_X'''$ & $\times$\\ \hline
$(7,2)$ & $\lambda, \kappa, \kappa', \lambda_X, \lambda_X',\lambda_X'', \lambda_X'''$ & $\times$\\ \hline
$(7,3)$ & $\lambda, \kappa, \kappa', \lambda_X, \lambda_X',\lambda_X'', \lambda_X'''$ & $\times$\\ \hline
\end{tabular}
\caption{The independent coupling constants in the scalar potential are listed. %\red{need not to consider (5,0) and (7,0)?}
We also show the dimension four extra couplings which contribute to the beta functions. 
}
\label{Table:coupling}
%\end{center}
\end{table}
%++++++++++++++++++++++++++++++++++++++++++++++++++
~\\

%%%%%%%%%%%%%%%%%%%% RGE %%%%%%%%%%%%%%%%%%%%

Next, let us move to the RGE analysis. 
The scale below the mass of $X$, $M_X$, we use the SM beta functions. 
Above $M_X$, the runnings of the electroweak gauge couplings are modified to as
%----------------------------------------
\begin{align}
\frac{dg_Y^{}}{dt}
&={1\over16\pi^2}
\left(\frac{41}{6}+\frac{1}{6}A\, Y_X^2\, n_X\right)g_Y^3, \\
\frac{dg_2}{dt}
&={1\over16\pi^2}
\left(-{19\over6}+{1\over6}A\, T(X)\right)g_2^3,
\end{align}
%----------------------------------------
where $t=\ln\mu$ with $\mu$ being the renormalization scale, 
$A=1\, (2)$ for a real (complex) scalar field, $T(X)$ is the Dynkin index. 
We see that $g_2$ is non-asymptotic free if $n_X^{} \geq 7\, (5)$ for a real (complex) field case. 
The runnings of the top-Yukawa coupling $y_t$ and the strong gauge coupling $g_3$ are unchanged at one-loop level. 
The one-loop RGEs of the scalar coupling constants are given in Appendix~\ref{sec:RGE}. 
~\\

%%%%%%%%%%%%%%%%%%%% INITIAL CONDITIONS %%%%%%%%%%%%%%%%%%%%

As a set of initial conditions for the RGEs, we take
%----------------------------------------
\begin{align}
\label{initial condition}
\lambda_X(M_X)=\lambda_X'(M_X)=\cdots=\kappa(M_X)=\kappa'(M_X)=0,
\end{align}
%----------------------------------------
and evaluate the scale of the LP. 
This set gives a conservative evaluation for the scale of the LP  
under the condition that the scalar potential is bounded from below.
%
%\blue{The cases with non-zero values of the initial condition are also shown in the numerical analysis.}
%
We use the initial values of SM parameters in Ref.~\cite{Buttazzo:2013uya} and 
take $M_H=125.7\GeV, M_t=173\GeV$, which are not sensitive to our analysis. 
By calculating the RGE, the existence of the LP is found below Planck scale for $n_X\geq4$. 
The blowup of the quartic coupling constants $\lambda_X, \lambda_X', \cdots$ of $X$  occurs 
because the quantum corrections to these couplings are proportional to the fourth power of 
the electroweak charges of $X$. 
This gives a rapid injection of non-zero value to the quartic coupling constants,  
and eventually results in the LP.
~\\

\begin{figure}[tb]
\begin{center}
\hfill
\includegraphics[width=.4\textwidth]{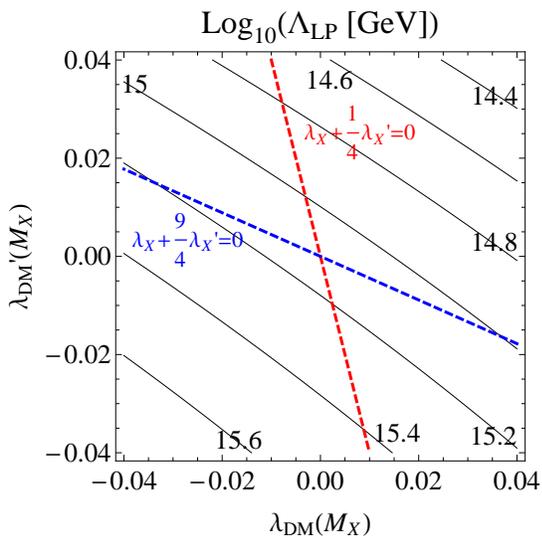}
%\hfill
%\includegraphics[width=.49\textwidth]{self.pdf}
\hfill\mbox{}\\
\caption{A contour plot of the scale of the LP as functions of $\lambda_X, \lambda_X'$ at $\mu=M_X$ with $M_X=100\GeV$.
The red and blue lines represent to $\lambda_X+1/4\lambda_X'=0$ and $\lambda_X+9/4\lambda_X'=0$, respectively. 
The left-bottom side of the red or blue line corresponds to the unbounded scalar potential. 
}
\label{Fig:4rep}
\end{center}
\end{figure}

% Discussion about initial consitions
%
Let us discuss the initial condition for the new quartic couplings constants. 
For an illustration, we focus on the case $(n_X^{}, Y_X)=(4, 3/2)$ 
with an additional $Z_2$ symmetry, %$X \to -X$, 
which forbids the term ${\Phi^\dag}^3X$. 
In this case, by imposing the boundedness of the scalar potential for $|X^1|^4$ and $|X^2|^4$ directions, 
we obtain the following conditions\footnote{
Note that this is not the sufficient condition for the boundedness but the necessary condition. }
%----------------------------------------
\begin{align}
&\lambda_X+{1\over4}\lambda_X'\geq 0,
&\lambda_X+{9\over4}\lambda_X'\geq 0.
\end{align}
%----------------------------------------
%
This requirements comes from the $|X^1|^4$ and $|X^2|^4$ terms in the scalar potential of $X$, 
%----------------------------------------
\begin{align}
&
\lambda_X|X^\dagger X|^2+\lambda_X'(X^\dagger T^a X)^2\nonumber\\
&
=\lambda_X(|X^1|^4+|X^2|^4)
+\lambda_X'\left({9\over4}|X^1|^4+{1\over4}|X^2|^4\right)+\cdots,
\end{align}
%----------------------------------------
where the second term in the r.h.s. of the equation results from 
$(X^\dagger T^3 X)^2$ with $T^3=\text{diag} (3/2,1/2,-1/2,-3/2)$. 
The same conditions are also come from $|X^3|^4$ and $|X^4|^4$ directions. 
The contour plot for the scale of the LPs together with the above two conditions
is shown in Fig.~\ref{Fig:4rep} as functions of $\lambda_X(M_X)$ and $\lambda_X'(M_X)$. 
%
%Since the LP is related to the running of $\lambda_X, \lambda_X'$, 
% we here assume that all other couplings are equal to zero for the moment.
One can see that $\lambda_X=\lambda_X'=0$ actually gives the {\it conservative scale} (the largest scale) of the LP. 
% from the figure. 
%
Even if we introduce nonzero $\kappa, \kappa', \lambda_{{\Phi^\dagger}^3 X}$, 
%and $ \lambda_{\Phi {\Phi^\dagger}^2 X}$ in $()$, 
these effects decrease the scale of the LP since these contributions to the beta functions 
of $\lambda_X$ and $\lambda_X'$ are always positive. 
For the $(n_X^{}, Y_X)=(4, 1/2)$ case, the similar arguments are applicable 
for $\lambda_X, \lambda_X'$ and $ \lambda_{\Phi {\Phi^\dagger}^2 X}$. 
Although the effect of $\lambda_{{\Phi^\dagger}^2 X^2}$ is unclear, 
we numerically confirmed that this coupling constant also leads to the lower scale of the LP. 
Thus, we conclude that Eqs.~\eqref{initial condition} are the initial condition for the conservative evaluation 
of the LP.  
The same discussions would be applied to higher isospin multiplets. %, which
We also confirmed these points numerically. 
The detailed and generalized analysis will be presented in our future publications~\cite{HKT2}.
~\\
%

%%%%%%%%%%%%%%%%%%%% RESULTS %%%%%%%%%%%%%%%%%%%%
\begin{figure*}[tb]
\includegraphics[width=.41\textwidth]{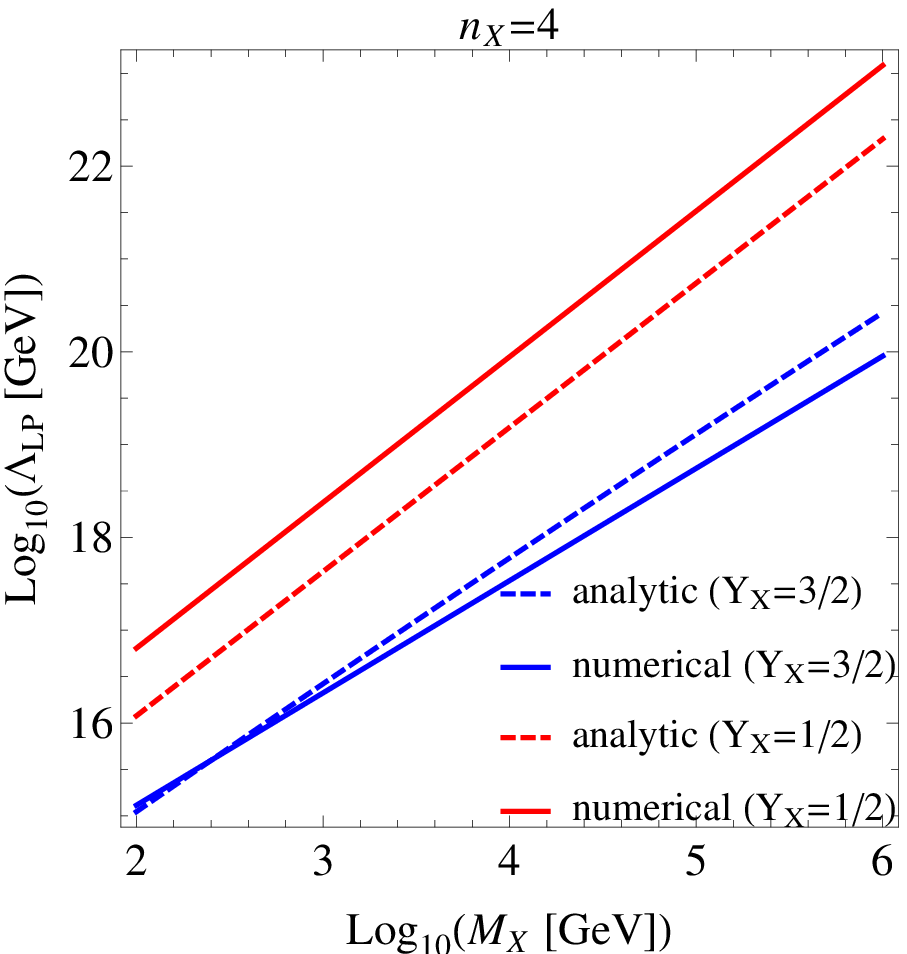}\hspace{10mm}
\includegraphics[width=.41\textwidth]{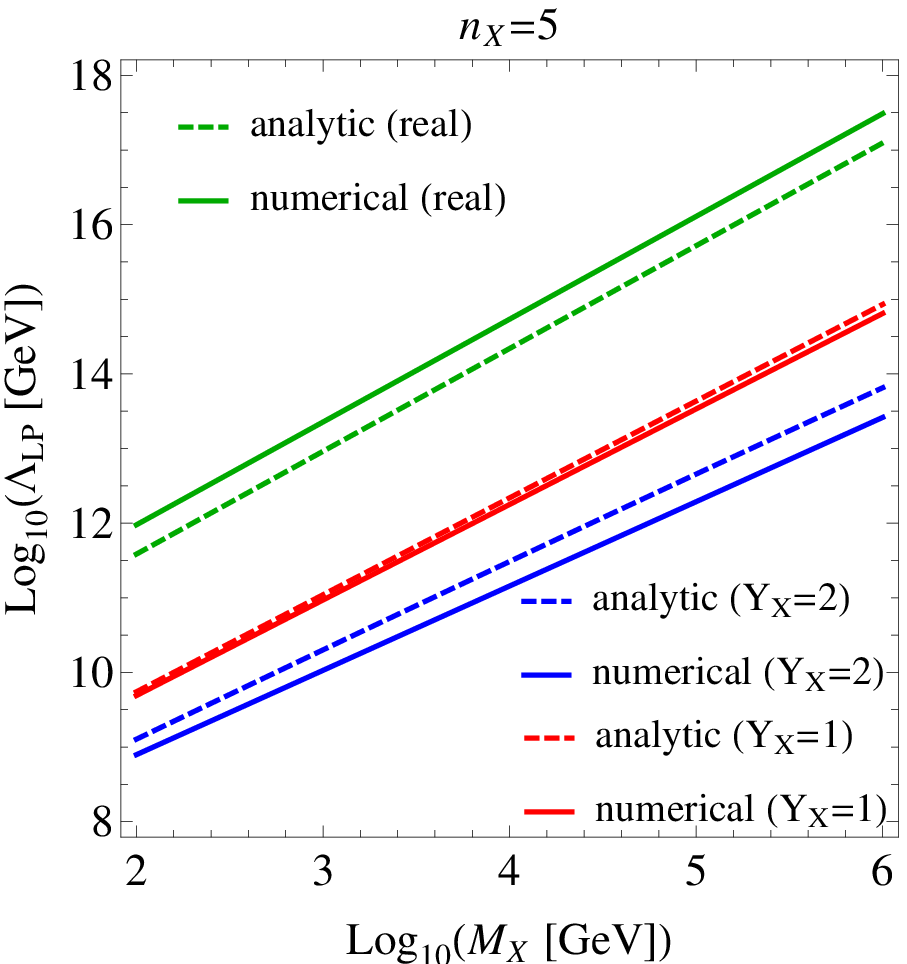}\\
\includegraphics[width=.41\textwidth]{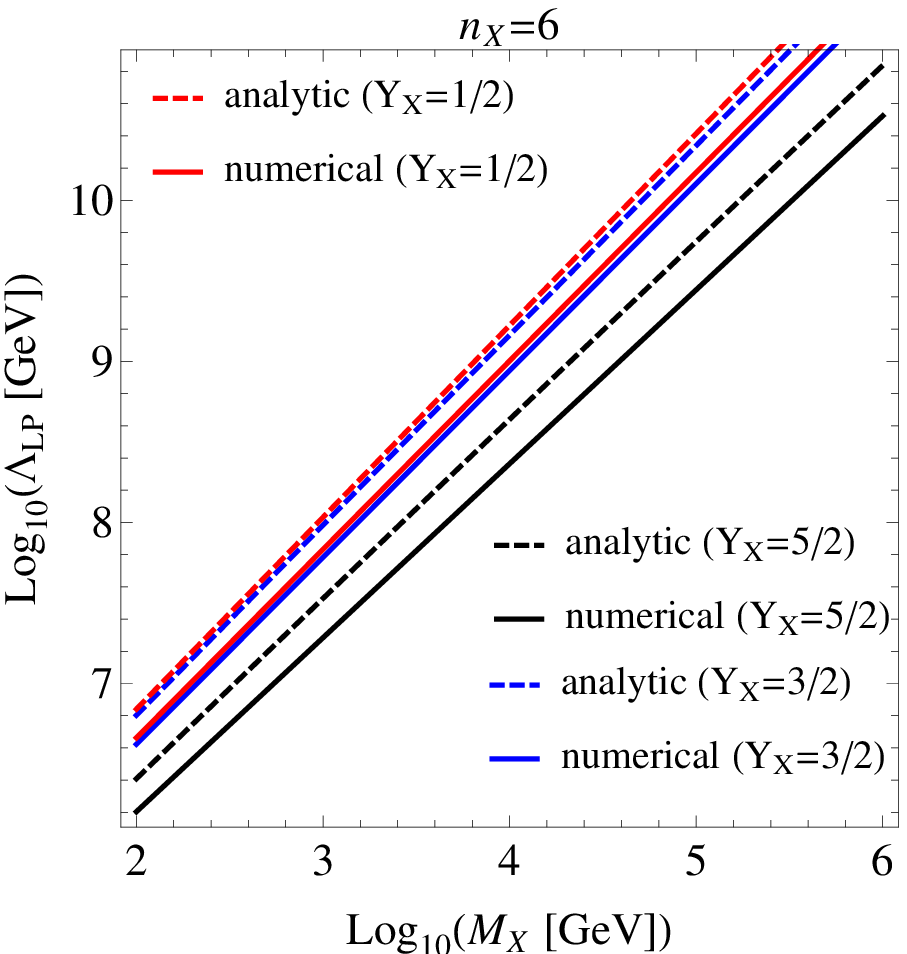}\hspace{12mm}
\includegraphics[width=.40\textwidth]{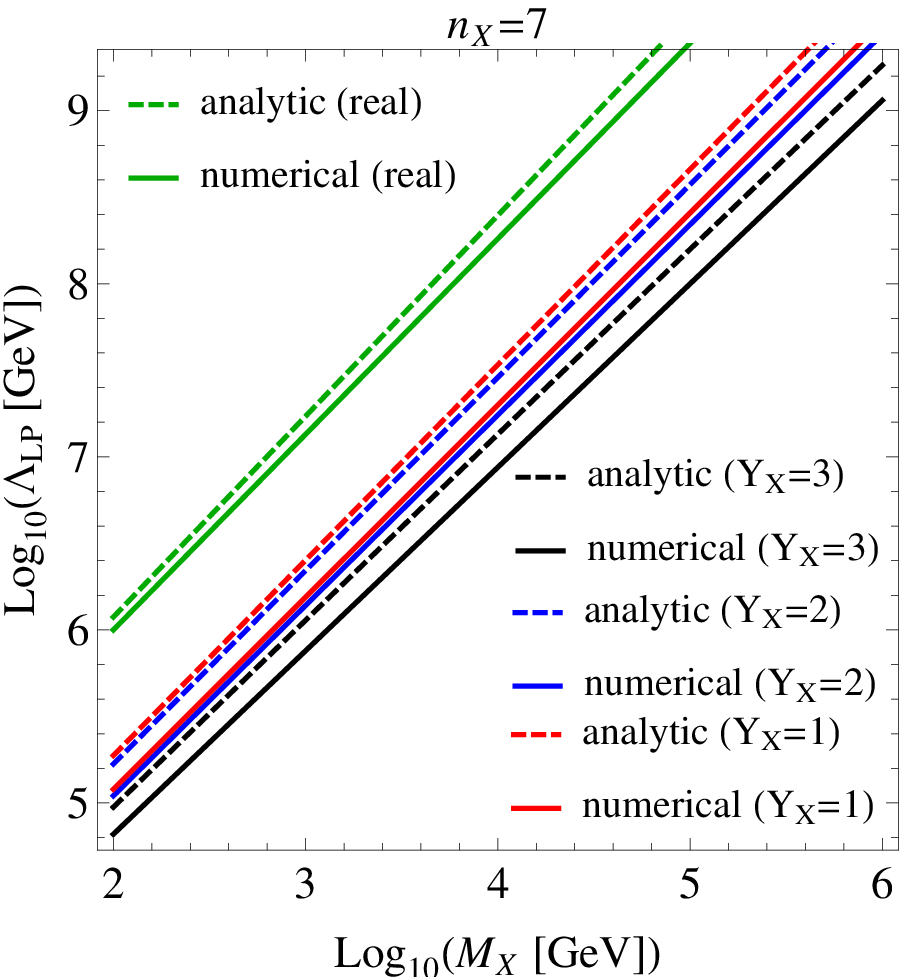}
\caption{
The solid and dashed lines correspond to the numerical results and the approximated results in 
Eq.~\eqref{analytic formula}.
}
\label{comparison}
\end{figure*} 
In Fig.~\ref{comparison}, we present the conservative scale of the LP as a function of $M_X$ 
in the SM with one more scalar multiplet $(n_X^{} = 4, 5, 6, 7)$.\footnote{
Requiring the tree-level unitarity of $SU(2)_L$ gauge interactions, 
$n_X^{} \le 8 (9)$ is obtained for a complex (real) scalar multiplet\cite{Hally:2012pu,Earl:2013jsa}.} 
We consider all the possible assignments of the hypercharges, where fractional electric charges are not allowed for new particles.
%From the top to the bottom, the results in the model with 
%a quadruplet $(Y_X=1/2, 3/2)$, a quintet $(Y_X=0, 1, 2)$, 
%a sextet $(Y_X=1/2, 3/2, 5/2)$, and septet $(Y_X=0, 1, 2, 3)$, where 
%$X^\dag = X$
For a $Y_X=0$ field, a real scalar condition is assumed. 
%a real scalar condition is assumed for a field with $Y_X=0$. 
%
The solid lines in each plot represent the numerical results, 
while the dashed lines show the approximated results obtained 
from the analytic study as we will discussed later in this letter.  
We fit the solid lines in Fig.~\ref{comparison} by
the scale of the LP and $M_X$ with an exponent.  
The results are listed in Table~\ref{Table:fitting}.
For comparisons, we also present the positions of the LP $\Lambda_{g_i}$ 
which are calculated by solving the one-loop beta function of gauge couplings.
%----------------------------------------
\begin{align}
\Lambda_{g_i}
&=M_X \exp 
%\left( {8\pi^2\over B_i}{1\over g_i^2(M_X)} \right)
\left( {1\over 2B_i}{1\over g_i^2(M_X)} \right) \nonumber \\
&=
M_X \left( {M_t\over M_X} \right)^{B_{i,\text{SM}}/B_i}
\exp \left( {1\over 2B_i}{1\over g_i^2(M_t)} \right),
\end{align}
%----------------------------------------
where $i=Y, 2$, and $B_{i,(\text{SM})}$ is the coefficient of the beta function of $g_i$ 
including a factor $1/16\pi^2$. 
%\begin{align}
%B_{i,(\text{SM})}=16\pi^2 {dg_{i,(\text{SM})}\over dt}%\beta_{g_{i,(\text{SM})}}.
%}
We see that the conservative scale of the LP from the scalar quartic coupling constant
is much lower than $\Lambda_{g_i}$. % from gauge coupling constant. 
%
%Furthermore, the scalar with $n_X\geq4$ introduces the pole below Planck scale.
%In this sense, $n_X\leq3$ is favored.
~\\

%
%++++++++++++++++++++++++++++++++++++++++++++++++++
\begin{table}[tb]
%\begin{center}
\begin{tabular}{|c|c|c|c|}
\hline
$(n_X^{},Y_X)$ & LP $[\GeV]$ & $\Lambda_{g_Y^{}}\, [\GeV]$ & $\Lambda_{g_2}\, [\GeV]$\\ \hline
$(4,1/2)$ & $6.4\times 10^{16}\left(M_X/100\GeV\right)^{1.57}$ & $1.2\times 10^{41}$ & -- \\ \hline
$(4,3/2)$ & $1.3\times 10^{15}\left(M_X/100\GeV\right)^{1.21}$ & $3.1\times 10^{30}$ & -- \\ \hline
$(5,\text{real})$ & $9.5\times 10^{11}\left(M_X/100\GeV\right)^{1.38}$ & $9.6\times 10^{42}$ & --\\ \hline
%$(5,0)$ & $4.9\times 10^{9}\left(M_X/100\GeV\right)^{1.2863}$ & $9.6\times 10^{42}$ & $9.0\times 10^{488}$\\ \hline
$(5,1)$ & $4.9\times 10^{9}\left(M_X/100\GeV\right)^{1.28}$ & $9.0\times 10^{34}$ & $9.0\times 10^{488}$\\ \hline
$(5,2)$ & $7.9\times 10^{8}\left(M_X/100\GeV\right)^{1.13}$ & $5.7\times 10^{22}$ & $9.0\times 10^{488}$\\ \hline
$(6,1/2)$ & $4.6\times 10^{6}\left(M_X/100\GeV\right)^{1.17}$ & $1.6\times 10^{40}$ & $2.7\times 10^{32}$\\ \hline
$(6,3/2)$ & $4.2\times 10^{6}\left(M_X/100\GeV\right)^{1.16}$ & $5.2\times 10^{26}$ & $2.7\times 10^{32}$\\ \hline
$(6,5/2)$ & $1.6\times 10^{6}\left(M_X/100\GeV\right)^{1.08}$ & $3.2\times 10^{16}$ & $2.7\times 10^{32}$\\ \hline
$(7,\text{real})$ & $1.0\times 10^{6}\left(M_X/100\GeV\right)^{1.13}$ & $9.6\times 10^{42}$ & $1.3\times 10^{56}$\\ \hline
%$(7,0)$&$1.2\times 10^{5}\left(M_X/100\GeV\right)^{1.10817}$ & $9.6\times 10^{42}$ & $1.4\times 10^{15}$\\ \hline
$(7,1)$ & $1.2\times 10^{5}\left(M_X/100\GeV\right)^{1.11}$ & $3.6\times 10^{32}$ & $1.4\times 10^{15}$\\ \hline
$(7,2)$ & $1.1\times 10^{5}\left(M_X/100\GeV\right)^{1.10}$ & $2.2\times 10^{19}$ & $1.4\times 10^{15}$\\ \hline
$(7,3)$ & $6.6\times 10^{4}\left(M_X/100\GeV\right)^{1.06}$ & $1.2\times 10^{12}$ & $1.4\times 10^{15}$\\ \hline
\end{tabular}
\caption{The fitted results of the scale of LPs with the initial condition in Eqs~\eqref{initial condition}. 
%This table gives the conservative values of pole.
The positions of LP derived from one-loop beta function of the gauge couplings taking $M_X=100\GeV$ 
are also given. 
%\red{need not to consider (5,0) and (7,0)?}
}
\label{Table:fitting}
%\end{center}
\end{table}
%++++++++++++++++++++++++++++++++++++++++++++++++++

%%%%%%%%%%%%%%%%%%%% DISCUSSION %%%%%%%%%%%%%%%%%%%%

% About lambda_EFF
%
In order to understand the results analytically, 
we focus on the beta function of the quartic coupling of $X$ 
in which the gauge coupling $g(\mu)$ is approximated by the constant $g(M_X)$. 
Then, we redefine the quartic coupling constants 
$\lambda_X', \cdots$ such that $g^4$ terms only appear in one beta function. 
By this redefinition, we correctly take into account the induced quartic coupling constant 
from the gauge couplings at one-loop level, since we begin with a conservative assumption 
given in Eqs.~\eqref{initial condition}.
%\footnote{This approach is useful only for the case where the electroweak charges of $X$ is relatively small.  
%If the charges are large, .....}
%
We call the new coupling constants $\tilde{\lambda}_X', \cdots$. 
%among which only the beta function of $\lambda_X$ contains $g^4$ term. 
%
Consequently, the RGE is simplified as
%----------------------------------------
\begin{align}
\frac{d\lambda_X}{dt}	
&=   {1\over16\pi^2}(c_0-c_1 \lambda_X+c_2 \lambda_X^2
+c_3\lambda_X\tilde{\lambda}_X'+c_4\tilde{\lambda}_X'^2), 
\\
\frac{d\tilde{\lambda}_X'}{dt}
&
=   {1\over16\pi^2}(-\tilde{c}_1' \lambda_X+\tilde{c}_2' \lambda_X^2
+\tilde{c}_3'\lambda_X\tilde{\lambda}_X'+\tilde{c}_4'\tilde{\lambda}_X'^2). 
\end{align}
%----------------------------------------
However, even if we set $\tilde{\lambda}_X'=0$ at $M_X$, nonzero $\tilde{\lambda}_X'$ is induced by the $\tilde{c}_2'$ term, which affects the running of $\lambda_X$ through the $c_3$ term.\footnote{
We checked that the effect of the $c_4$ term is smaller than that of the $c_3$ term. 
%\red{Can we argue that the effect comes from 2-loop?}
}
Therefore, we redefine $\lambda_X$ in order to cancel the $c_3$ term.
As a result, we have 
\begin{align}
\label{redefined rge}
\frac{d\tilde{\lambda}_X}{dt}	
&=   {1\over16\pi^2}(\tilde{c}_0-\tilde{c}_1 \tilde{\lambda}_X+\tilde{c}_2 \tilde{\lambda}_X^2),
\end{align}
taking $\tilde{\lambda}_X'= \cdots =0$. 
This simplified differential equation can be solved analytically. 
The position of the LP is calculated as 
%----------------------------------------
\begin{align}
\label{analytic formula}
\frac{\Lambda_{\text{LP}}}{M_X}
&=\exp\Bigg[
{16\pi^2\over \tilde{c}_2d}
\bigg\{\frac{\pi}2 -
\tan^{-1} \bigg[ \frac1{d} 
\bigg(\tilde{\lambda}_X(M_X)-{\tilde{c}_1\over2\tilde{c}_2}\bigg)
\bigg] \bigg\} \Bigg], \\
d&=\sqrt{{\tilde{c}_0\over \tilde{c}_2}-{\tilde{c}_1^2\over4\tilde{c}_2^2}}.
\end{align}
%----------------------------------------
%
As an example, we show the concrete procedure for the case with $n_X=4$. %$(n_X^{},Y_X)=(4, 3/2)$.  
Concentrating on the quartic coupling constants of $X$, we have
%----------------------------------------
\begin{align}
&\frac{d\lambda_X}{dt} = {1\over16\pi^2} \bigg(
+32\lambda_X^2+30\lambda_X\lambda_X'+{99\over2}\lambda_X'^2 \nonumber \\
&\qquad
+6Y_X^4g_Y^4 +{297\over8}g_2^4 -12Y_X^2\lambda_X g_Y^2 -45\lambda_X g_2^2
%\nonumber \\
%&\qquad
%+2\kappa^2 +2|\lambda_{{\Phi^\dagger}^2X^2}|^2
\bigg),\\
%==========
&\frac{d\lambda_X'}{dt} = {1\over16\pi^2} \bigg(
+24\lambda_X\lambda_X' +42\lambda_X'^2 +12Y_X^2g_Y^2 g_2^2 \nonumber \\
&\qquad
-12Y_X^2\lambda_X'g_Y^2 -45\lambda_X'g_2^2 
%+{1\over2}\kappa'^2 -{8\over9}|\lambda_{{\Phi^\dagger}^2X^2}|^2 
\bigg).
\end{align}
%----------------------------------------
Here we put $g_Y^{}(\mu)=g_Y^{}(M_X), g_2(\mu)=g_2(M_X)$ and
$\kappa=\kappa'=\lambda_{\Phi{\Phi^\dagger}^2X}=\lambda_{{\Phi^\dagger}^2X^2}
=\lambda_{{\Phi^\dagger}^3X}=\lambda_{{\Phi^\dagger}^3X}=0$. 
At first, we redefine $\lambda_X'$ in order to cancel the constant term ($12 Y_X^2 g_Y^2 g_2^2$ term).
Hence, we take
%----------------------------------------
\begin{align}
\tilde{\lambda}_X' =\lambda_X' +\eta\, \lambda_X,
\end{align}
%----------------------------------------
where $\eta \equiv -12Y_X^2 g_Y^2 g_2^2/(6Y_X^4g_Y^4+{297\over8}g_2^4)$. 
%and the resultant beta functions are
%%----------------------------------------
%\begin{align}
%&\frac{d\lambda_X}{dt} = {1\over16\pi^2} \bigg\{
%\left(32-30F+{99\over2}F^2\right)\lambda_X^2
%+(30-99F)\lambda_X\tilde{\lambda}_X' \nonumber \\
%&\qquad
%+{99\over2}\tilde{\lambda}_X'^2
%+6Y_X^4g_Y^4 +{297\over8}g_2^4 -12Y_X^2\lambda_X g_Y^2 -45\lambda_X g_2^2
%%\nonumber \\
%%&\qquad
%%+2\kappa^2 +2|\lambda_{{\Phi^\dagger}^2X^2}|^2
%\bigg\},\nonumber \\
%%==========
%&\frac{d\tilde{\lambda}_X'}{dt} = {1\over16\pi^2} \bigg(
%(-24F+42F^2)\lambda_X^2
%+(24-84F)\lambda_X\tilde{\lambda}_X'  \nonumber \\ 
%&\qquad
%+42\tilde{\lambda}_X'^2
%-12Y_X^2\tilde{\lambda}_X'g_Y^2 -45\tilde{\lambda}_X'g_2^2 
%%+{1\over2}\kappa'^2 -{8\over9}|\lambda_{{\Phi^\dagger}^2X^2}|^2 
%\bigg),
%\end{align}
%%----------------------------------------
%where $F=-12Y_X^2 g_Y^2 g_2^2/(6Y_X^4g_Y^4+{297\over8}g_2^4)$.
%
The requirement of the vanishing $g_Y^2 g_2^2$ term fixes $\tilde{\lambda}_X'$
up to the normalization,  
while we still have the degree of freedom to redefine $\lambda_X$.
Using this, we define $\tilde{\lambda}_X$ as
%----------------------------------------
\begin{align}
\tilde{\lambda}_X=\lambda_X+\xi\, \tilde{\lambda}_X',
\end{align}
%----------------------------------------
and choose $\xi$ in such a way that the $\tilde{\lambda}_X \tilde{\lambda}_X'$ term vanishes 
in the beta function of $\tilde{\lambda}_X$. 
%%----------------------------------------
%\begin{align}
%&\frac{d\tilde{\lambda}_X}{dt} = {1\over16\pi^2} \bigg\{
%\left(32-30F+{99\over2}F^2-24FG+42F^2G\right)\tilde{\lambda}_X^2 \nonumber \\
%&\qquad
%+(30-99F-40G-24FG-99F^2G \nonumber \\
%&\qquad
%+48FG^2-84F^2G^2)\tilde{\lambda}_X\tilde{\lambda}_X' \nonumber \\
%&\qquad
%+({99\over2}+12G+99FG+8G^2+54FG^2 \nonumber \\
%&\qquad
%+{99\over2}F^2G^2-24F G^3+42F^2G^3)\tilde{\lambda}_X'^2\nonumber \\
%&\qquad
%+6Y_X^4g_Y^4 +{297\over8}g_2^4 -12Y_X^2\lambda_X g_Y^2 -45\lambda_X g_2^2
%%\nonumber \\
%%&\qquad
%%+2\kappa^2 +2|\lambda_{{\Phi^\dagger}^2X^2}|^2
%\bigg\},\nonumber \\
%%==========
%&\frac{d\tilde{\lambda}_X'}{dt} = {1\over16\pi^2} \bigg(
%(-24F+42F^2)\tilde{\lambda}_X^2  \nonumber \\
%&\qquad
%+(24-84F+48FG-84F^2G)\tilde{\lambda}_X\tilde{\lambda}_X'  \nonumber \\ 
%&\qquad
%+(42-24G+84FG-24FG^2+42F^2G^2)\tilde{\lambda}_X'^2  \nonumber \\
%&\qquad
%-12Y_X^2\tilde{\lambda}_X'g_Y^2 -45\tilde{\lambda}_X'g_2^2 
%%+{1\over2}\kappa'^2 -{8\over9}|\lambda_{{\Phi^\dagger}^2X^2}|^2 
%\bigg),
%\end{align}
%%----------------------------------------
Then, we obtain
%----------------------------------------
\begin{align}
\xi={-20+3\eta-99\eta^2\over16\eta+24\eta^2+99\eta^3}%\nonumber\\
%&\qquad
\pm{\sqrt{5(80+72\eta+621\eta^2)}
\over16\eta+24\eta^2+99\eta^3}.
\end{align}
%----------------------------------------
%$\tilde{\lambda}_X= ...$ and $\tilde{\lambda}'_X= ...$
%for the case with $(n_X^{},Y_X)=(4, 3/2)$. 
The coefficients $c_0, c_1, c_2$ are given by
%----------------------------------------
\begin{align}
&\tilde{c}_0={297\over8}g_2^4+6Y_X^4g_Y^4,\\
&\tilde{c}_1=12Y_X^2g_Y^2+45g_2^2,\\
&\tilde{c}_2=32-30\eta+{99\over2}\eta^2+8\eta\xi+12\eta^2\xi+{99\over2}\eta^3\xi. 
\end{align}
%----------------------------------------
%
By this procedure, we can evaluate the LP in the analytical calculations.  
The approximated lines are also plotted in Fig.~\ref{comparison}. 
Comparing the numerical results and our analytic results in these figures, 
the above expressions fit the numerical results within an order of magnitude error. 
From this analytical analysis, we confirm that the finite quartic coupling constants are originally 
generated by the gauge couplings, and then the RGE of the quartic coupling constant leads the LP. 
Roughly speaking, the LP is induced by two-loop radiative corrections to the RGEs. 
We can also evaluate the exponent of $M_X$ 
by putting $g_{Y(2)}(M_X)\simeq g_{Y(2)}(M_t)+B_{Y(2),\text{SM}}\ln (M_X/M_t)$.
Although the analytic expression is too complicate to write,
we obtain $\sim1.52$ and $\sim1.28$ for $(4,1/2)$ and $(4,3/2)$, respectively. 
~\\

%%%%%%%%%%%%%%%%%%%% REMARKS %%%%%%%%%%%%%%%%%%%%
Before concluding this letter, a few more remarks are given in order. 
Firstly, the position of the LP does not depend on the normalization of $\tilde{\lambda}_X$. 
We can understand the reason from Eqs.~\eqref{redefined rge} and \eqref{analytic formula} 
as follows; let us define the new scalar coupling constant having a different normalization 
as $\tilde{\lambda}^{N}_{X}:=a\,\tilde{\lambda}_{X}$ where $a$ is a real constant. 
Although the RGE of $\tilde{\lambda}^{N}_{X}$ takes the same form as $\tilde{\lambda}_{X}$, 
the coefficients are different. They are $a\, c_{0}$, $c_{1}$ and $c_{2}/a$. 
By using these coefficients and $\tilde{\lambda}^{N}_{X}(M_{X})=a\, \tilde{\lambda}_{X}(M_{X})$, 
one can easily show that the LP for $\tilde{\lambda}^{N}_{X}$ calculated by Eq.~\eqref{analytic formula} 
gives the same result as that of $\tilde{\lambda}_{X}$. Therefore, we do not need to care about 
the normalizations of scalar couplings when we study the LP. 

Secondly, we discuss the nonzero value of the initial condition.
Let us start with Eq.~\eqref{analytic formula}.
Expanding Eq.~\eqref{analytic formula} around $1/\tilde{\lambda}_X(M_{X})=0$, we get
%----------------------------------------
\begin{align}
{\Lambda_\text{LP}\over M_X}
=\exp\Bigg[
{16\pi^2\over \tilde{c}_2 }{1\over \tilde{\lambda}_X(M_{X})}
+\mathcal{O}\Big(1/\tilde{\lambda}_X(M_{X})^2\Big)
\Bigg],
\end{align}
%----------------------------------------
which means that the scale of the LP is determined by only the initial value and $\tilde{c}_2$.
The conditions to neglect higher order terms are $\tilde{\lambda}_X(M_{X})\gtrsim \tilde{c}_1/(2\tilde{c}_2), d$. 
Numerically, $\tilde{c}_1/(2\tilde{c}_2) \approx 0.28, d \approx 0.31$ for $(n_X^{}, Y_X)=(4, 3/2)$.

Though we have focused on a single scalar extension of the SM in this letter, 
more matter fields may be added.  
Since the increase of scalar fields leads to the large coefficient in the beta functions in general, 
we expect that the scale of the LP becomes lower. 
If we add fermions in addition to scalars, the Yukawa couplings among them may be allowed.
In such a case, 
the scale of the LP can be large
%(fermion case, including right-handed neutrino Yukawa)
since the Yukawa coupling gives the negative contribution to the beta function at the one-loop level.
%
%
%In supersymmetric models, we should introduce at least two scalars for $Y_X\neq0$.
%
%Furthermore, for $Y_X\neq0(\text{not necessary??})$ and $n_X\geq4$, 
%there are no dimension three couplings in the superpotential except for 
%the SM ones(correct?), and so all the scalar couplings are determined 
%by the gauge coupling. 
%In this sense, our argument can not apply to the supersymmteric models.

Finally, we comment on the implications of the new physics beyond the SM. 
In the Minimal dark matter models~\cite{Cirelli:2005uq,Cirelli:2007xd,Cirelli:2009uv}, 
interactions of new scalar multiplets are also bounded by the relic abundance 
of dark matter and the data for direct/indirect searches. 
It is very interesting to combine our study with such constraints\cite{HKT2}. 
In some class of the seesaw models for the neutrino mass generation, 
Majorana fermions are introduced together with scalar multiplets, which may or may not  
accommodate a dark matter candidate. These models are constrained not only 
by the relic abundance but also by the neutrino oscillation data. 
In these models, the effects of Majorana fermions on the RGEs are expected to be small, 
because the required Yukawa coupling is generally small if we add these new fermions in the electroweak/TeV scale. 
In addition,
models with a flavor symmetry are also very intriguing candidates for this kind of analyses. 
Because the scalar fields are embedded in the multiplet under the flavor symmetry, 
which provides a large number of $SU(2)_L$ multiplets. Increase of the number of 
multiplet also gives large contribution to the RGE, which would make the LP smaller\cite{HKT2}. 
~\\

%%%%%%%%%%%%%%%%%%%% CONCLUSION %%%%%%%%%%%%%%%%%%%%
In conclusion, we have investigated the scale of the LP 
using the one-loop RGE in the SM with one more scalar $n_X^{}$-plet $X$
($n_X^{}=4, 5, 6, 7$ with all possible hypercharge assignments). 
The LP is found below Planck scale for $n_X^{} \ge 4$, 
if we introduce a new scalar multiplet at electroweak/TeV scale. 
This means that any single scalar field extension with $n_X^{} \ge 4$ of the SM is not allowed, if we impose the absence of the LP up to Planck scale.
The scale is evaluated with the conservative initial condition, where 
the initial values for the quartic coupling constant of new scalar fields 
are set to be zeros at the scale of new particles. 
Nevertheless, the quartic coupling constants are induced 
from the electroweak gauge interactions. 
The induced coupling constants are rapidly enhanced, and finally hit the LP. 
We have calculated the LP as a function of the mass of $X$, and 
listed a fitting formula of the LP with an exponent. 
The results are consistent with our approximated formula within an order of magnitude error, 
which is derived from the simplified RGE with the conservative assumption. 
The obtained LP in each model is much smaller than that previously calculated 
by solving the beta functions of the gauge couplings\cite{DiLuzio:2015oha}. 
These new results are very useful and generic constraints on the beyond the SM 
including new scalar multiplets.

%%%%%%%%%%%%%%%%%%%% ACKNOWLEDGMENTS %%%%%%%%%%%%%%%%%%%%
\section*{Acknowledgments}\vspace{-3ex}
This work is supported by the Grant-in-Aid for Japan Society for the Promotion of Science (JSPS) Fellows 
No.25$\cdot$1107 (YH) and No.27$\cdot$1771 (KK). 
K.T.'s work is supported in part by the MEXT Grant-in-Aid for Scientific Research on Innovative Areas No. 26104704. 
\\

%%%%%%%%%%%%%%%%%%%% APPENDIX %%%%%%%%%%%%%%%%%%%%
\appendix
\section{Renormalization Group equations}\label{sec:RGE}
We calculate the RGEs by using the general formula in Refs.~\cite{Cheng:1973nv}.

\begin{itemize}
%\item real triplet
%\begin{align}
%16\pi^2\frac{d\lambda}{dt}	
%        &=   
%                                   \bigg({3\over2}\kappa^2+12\lambda y_t^2+24\lambda^2-6y_t^4-9\lambda g_2^2+{9\over8}g_2^4
%                                          -3\lambda g_Y^2+{3\over4}g_Y^2g_2^2+{3\over8}g_Y^4
%                                   \bigg),\nonumber \\
%16\pi^2\frac{d\lambda_X}{dt}	
%        &=   {1\over16\pi^2}
%                                   \bigg(
%                                   2\kappa^2+22\lambda_X^2                                   
%                                   -24\lambda_X g_2^2+12g_2^4                                   %&-12Y_X^2\lambda_X g_Y^2-9\left({C_2(S_X)\over C_2(\Phi)}\right)\lambda_X g_2^2
%                                   \bigg),\nonumber \\
%16\pi^2\frac{d\kappa}{dt}	
%        &=   {1\over16\pi^2}
%                                   \bigg(6y_t^2\kappa+4\kappa^2
%                                          +12\kappa\lambda+10\kappa\lambda_X-{33\over2}\kappa g_2^2+6g_2^4-{3\over2}\kappa g_Y^2\bigg).
%}
%\newpage
%
%++++++++++++++++++++++++++++++++++++++++++++++++++++++++++++++++++++++
\item SM with a quadruplet scalar field
%$\lambda_{\Phi^2 X^2}$ and $\lambda_{\Phi^\dagger \Phi^2 X}$ exist only if $Y_X=-1/2$, and
%$\lambda_{\Phi^3 X}$ exists only if $Y_X=-3/2$.
\begin{align}
%==========
&\frac{d\lambda}{dt} = {1\over16\pi^2} \bigg(
+24\lambda^2  -6y_t^4 
+{3\over8}g_Y^4 +{9\over8}g_2^4 +{3\over4}g_Y^2g_2^2 \nonumber \\
&\qquad 
+12\lambda y_t^2 -3\lambda g_Y^2 -9\lambda g_2^2 
+4\kappa^2 +{5\over4}\kappa'^2 \nonumber \\
&\qquad 
+{40\over9}|\lambda_{{\Phi^\dagger}^2X^2}|^2 +8|\lambda_{\Phi {\Phi^\dagger}^2 X}|^2
+24\lambda_{{\Phi^\dagger}^3 X}^2
\bigg),\nonumber \\
%==========
&\frac{d\lambda_X}{dt} = {1\over16\pi^2} \bigg(
+32\lambda_X^2+30\lambda_X\lambda_X'+{99\over2}\lambda_X'^2 \nonumber \\
&\qquad
+6Y_X^4g_Y^4 +{297\over8}g_2^4 -12Y_X^2\lambda_X g_Y^2 -45\lambda_X g_2^2\nonumber \\
&\qquad
+2\kappa^2 +2|\lambda_{{\Phi^\dagger}^2X^2}|^2
\bigg),\nonumber \\
%==========
&\frac{d\lambda_X'}{dt} = {1\over16\pi^2} \bigg(
+24\lambda_X\lambda_X' +42\lambda_X'^2 +12Y_X^2g_Y^2 g_2^2 \nonumber \\
&\qquad
-12Y_X^2\lambda_X'g_Y^2 -45\lambda_X'g_2^2 +{1\over2}\kappa'^2 -{8\over9}|\lambda_{{\Phi^\dagger}^2X^2}|^2 
\bigg),\nonumber \\
%==========
&\frac{d\kappa}{dt} = {1\over16\pi^2} \bigg(
+3Y_X^2 g_Y^4 +{45\over4}g_2^4 +4\kappa^2+{15\over4}\kappa'^2 \nonumber \\
&\qquad
+6y_t^2\kappa -6Y_X^2\kappa g_Y^2 -{3\over2}\kappa g_Y^2 
-27\kappa g_2^2 \nonumber \\
&\qquad
+12\kappa\lambda
+20\kappa\lambda_X+15\kappa\lambda_X' \nonumber \\
&\qquad
+18\lambda_{{\Phi^\dagger}^3 X}^2
+{40\over3}|\lambda_{{\Phi^\dagger}^2X^2}|^2+6|\lambda_{\Phi {\Phi^\dagger}^2 X}|^2
\bigg),\nonumber \\
%==========
&\frac{d\kappa'}{dt} = {1\over16\pi^2} \bigg(
+12Y_X g_Y^2 g_2^2 \nonumber \\
&\qquad
+6y_t^2\kappa' -{3\over2}\kappa'g_Y^2 -6Y_X^2\kappa'g_Y^2 -27\kappa'g_2^2\nonumber \\
&\qquad
+8\kappa\kappa' +4\kappa'\lambda +4\kappa'\lambda_X +31\kappa'\lambda_X'
\nonumber \\
&\qquad
+24\lambda_{{\Phi^\dagger}^3 X}^2 +{64\over9}|\lambda_{{\Phi^\dagger}^2X^2}|^2 
+{8\over3}|\lambda_{\Phi {\Phi^\dagger}^2 X}|^2
\bigg),\nonumber \\
%==========
&\frac{d\lambda_{{\Phi^\dagger}^2X^2}}{dt} = {1\over16\pi^2} \bigg(
-4\lambda_{\Phi {\Phi^\dagger}^2 X}^2 
+6y_t^2 \lambda_{{\Phi^\dagger}^2X^2} \nonumber \\
&\qquad
-{3\over2}\lambda_{{\Phi^\dagger}^2X^2} g_Y^2 
-6Y_X^2 \lambda_{{\Phi^\dagger}^2X^2} g_Y^2
-27 \lambda_{{\Phi^\dagger}^2X^2} g_2^2 \nonumber \\
&\qquad
+8\kappa \lambda_{{\Phi^\dagger}^2X^2}
+4\kappa'  \lambda_{{\Phi^\dagger}^2X^2}  \nonumber \\
&\qquad
+4 \lambda \lambda_{{\Phi^\dagger}^2X^2}
+4\lambda_X  \lambda_{{\Phi^\dagger}^2X^2}
-11\lambda_X'  \lambda_{{\Phi^\dagger}^2X^2}
%-{3\over2}y_t^2 \mathrm{Re}(\lambda_{{\Phi^\dagger}^2X^2})\nonumber \\
%& \phantom{aaaaaaaa} +{27\over4}g_2^2 \mathrm{Re}(\lambda_{{\Phi^\dagger}^2X^2})
%+{3\over8}g_Y^2 \mathrm{Re}(\lambda_{{\Phi^\dagger}^2X^2})+{3\over2}Y_X^2 g_Y^2  \mathrm{Re}(\lambda_{{\Phi^\dagger}^2X^2})
\bigg), \nonumber \\
%==========
&\frac{d\lambda_{\Phi {\Phi^\dagger}^2 X}}{dt} = {1\over16\pi^2} \bigg(
+12\lambda \lambda_{\Phi {\Phi^\dagger}^2 X}
+9y_t^2 \lambda_{\Phi{\Phi^\dagger}^2X} \nonumber \\
&\qquad
-{9\over4}g_Y^2\lambda_{\Phi{\Phi^\dagger}^2X}
-3Y_X^2 g_Y^2\lambda_{\Phi{\Phi^\dagger}^2X}
-18g_2^2\lambda_{\Phi{\Phi^\dagger}^2X}
\nonumber \\
&\qquad
+6\kappa \lambda_{\Phi {\Phi^\dagger}^2 X} +{5\over2}\kappa'\lambda_{\Phi {\Phi^\dagger}^2 X}
-{40\over3}\lambda_{{\Phi^\dagger}^2X^2}\lambda_{\Phi {\Phi^\dagger}^2 X}^*
\bigg),\nonumber \\
%==========
&\frac{d\lambda_{{\Phi^\dagger}^3 X}}{dt} = {\lambda_{{\Phi^\dagger}^3 X}\over16\pi^2} \bigg(
+12\lambda +9y_t^2 -18g_2^2-{9\over4}g_Y^2-3Y_X^2 g_Y^2 \nonumber \\
&\qquad
+6\kappa+{15\over2}\kappa'
\bigg).
%==========
\end{align}

%++++++++++++++++++++++++++++++++++++++++++++++++++++++++++++++++++++++
\item SM with a real quintet scalar field
\begin{align}
%==========
&\frac{d\lambda}{dt} = {1\over16\pi^2} \bigg(
+24\lambda^2-6y_t^4 +{3\over8}g_Y^4+{9\over8}g_2^4 +{3\over4}g_Y^2g_2^2 \nonumber \\
&\qquad 
+{5\over2}\kappa^2 +12\lambda y_t^2 -3\lambda g_Y^2 -9\lambda g_2^2
\bigg),\nonumber \\
%==========
&\frac{d\lambda_X}{dt} = {1\over16\pi^2} \bigg(
+26\lambda_X^2
+108g_2^4
%+48\lambda_X\lambda_X'
%+720\lambda_X\lambda_X''+1152\lambda_X'\lambda_X''
-72\lambda_X g_2^2%+\red{0\times g_2^4+0\times g_Y^2g_2^2}
+2\kappa^2
%+\red{0\times g_2^4+0\times g_Y^2g_2^2}\nonumber \\
%-9\left({C_2(S_X)\over C_2(\Phi)}\right)\lambda_X g_2^2
\bigg),\nonumber \\
%==========
&\frac{d\kappa}{dt} = {1\over16\pi^2} \bigg(
+18g_2^4
+12\kappa\lambda +14\kappa\lambda_X%+24\kappa\lambda_X'
+6y_t^2\kappa \nonumber \\
&\qquad 
-{3\over2}\kappa g_Y^2 -{81\over2}\kappa g_2^2 +4\kappa^2
\bigg).
%==========
\end{align}

%++++++++++++++++++++++++++++++++++++++++++++++++++++++++++++++++++++++
\item SM with a complex quintet scalar field
\begin{align}
%==========
&\frac{d\lambda}{dt} = {1\over16\pi^2} \bigg(
+24\lambda^2 -6y_t^4 +{3\over8}g_Y^4 +{9\over8}g_2^4 +{3\over4}g_Y^2g_2^2 \nonumber \\
&\qquad
+12\lambda y_t^2 -3\lambda g_Y^2 -9\lambda g_2^2
+5\kappa^2+{5\over2}\kappa'^2
\bigg),\nonumber \\
%==========
&\frac{d\lambda_X}{dt} = {1\over16\pi^2} \big(
+36\lambda_X^2 +48\lambda_X\lambda_X' \nonumber \\
&\qquad
+720\lambda_X\lambda_X'' +1152\lambda_X'\lambda_X'' +3168\lambda_X''^2\nonumber \\
&\qquad
+6Y_X^4g_Y^4%+\red{0\times g_2^4+0\times g_Y^2g_2^2}
%+\red{0\times g_2^4+0\times g_Y^2g_2^2}\nonumber \\
%-9\left({C_2(S_X)\over C_2(\Phi)}\right)\lambda_X g_2^2
-12Y_X^2\lambda_X g_Y^2 -72\lambda_X g_2^2
+2\kappa^2
\big),\nonumber \\
%==========
&\frac{d\lambda_X'}{dt} = {1\over16\pi^2} \bigg(
+24\lambda_X\lambda_X' +84\lambda_X'^2
+408\lambda_X'\lambda_X'' -84\lambda_X''^2 \nonumber \\
&\qquad
+3g_2^4 +12Y_X^2g_Y^2g_2^2 %\nonumber \\
%&\qquad
-12Y_X^2\lambda_X'g_Y^2 -72\lambda_X'g_2^2%\red{+0\times g_Y^4}
+{1\over2}\kappa'^2
\bigg),\nonumber \\
%==========
&\frac{d\lambda_X''}{dt} = {1\over16\pi^2} \big(
+8\lambda_X'^2 +24\lambda_X\lambda_X'' -32\lambda_X'\lambda_X'' +368\lambda_X''^2\nonumber \\
&\qquad
+6g_2^4 -12Y_X^2\lambda_X'' g_Y^2 -72\lambda_X''g_2^2
 %\red{+0\times g_Y^4+0\times g_Y^2g_2^2}
\big),\nonumber \\
%==========
&\frac{d\kappa}{dt} = {1\over16\pi^2} \bigg(
+3Y_X^2 g_Y^4 +18g_2^4 \nonumber \\
&\qquad
+12\kappa\lambda +24\kappa\lambda_X +24\kappa\lambda_X' +360\kappa\lambda_X''
+6y_t^2\kappa \nonumber \\
&\qquad
-{3\over2}\kappa g_Y^2 -6Y_X^2\kappa g_Y^2 -{81\over2}\kappa g_2^2
+4\kappa^2 +6\kappa'^2
\bigg),\nonumber \\
%==========
&\frac{d\kappa'}{dt} = {1\over16\pi^2} \bigg(
+12Y_Xg_Y^2g_2^2 \nonumber \\
&\qquad
+4\kappa'\lambda +4\kappa'\lambda_X +60\kappa'\lambda_X' 
+60\kappa'\lambda_X'' +6y_t^2\kappa' \nonumber \\
&\qquad
-{3\over2}\kappa'g_Y^2 -6Y_X^2\kappa'g_Y^2 -{81\over2}\kappa'g_2^2
+8\kappa\kappa'
\bigg).
%==========
\end{align}

%++++++++++++++++++++++++++++++++++++++++++++++++++++++++++++++++++++++
\item SM with a sextet scalar field
\begin{align}
%==========
&\frac{d\lambda}{dt} = {1\over16\pi^2} \bigg(
+24\lambda^2-6y_t^4 +{3\over8}g_Y^4 +{9\over8}g_2^4 \nonumber \\
&\qquad
+12\lambda y_t^2 -3\lambda g_Y^2 -9\lambda g_2^2 +{3\over4}g_Y^2g_2^2 \nonumber \\
&\qquad
+6\kappa^2 +{35\over8}\kappa'^2 +{28\over5}|\lambda_{{\Phi^\dagger}^2X^2}|^2
\bigg),\nonumber \\
%==========
&\frac{d\lambda_X}{dt} = {1\over16\pi^2} \bigg(
+40\lambda_X^2+70\lambda_X\lambda_X'
+{3535\over2}\lambda_X\lambda_X''\nonumber \\
&\qquad
+{8525\over4}\lambda_X'\lambda_X''
+{271975\over16}\lambda_X''^2
 +6Y_X^4g_Y^4 \nonumber \\
&\qquad
-12Y_X^2\lambda_X g_Y^2 -105\lambda_X g_2^2 +2\kappa^2
-{11\over8}|\lambda_{{\Phi^\dagger}^2X^2}|^2
%+\red{0\times g_2^4+0\times g_Y^2g_2^2}
%+\red{0\times g_2^4+0\times g_Y^2g_2^2}\nonumber \\
%-9\left({C_2(S_X)\over C_2(\Phi)}\right)\lambda_X g_2^2
\bigg),\nonumber \\
%==========
&\frac{d\lambda_X'}{dt} = {1\over16\pi^2} \bigg(
+24\lambda_X\lambda_X'+136\lambda_X'^2 +{2115\over2}\lambda_X'\lambda_X'' \nonumber \\
&\qquad
-{265\over2}\lambda_X''^2 +3g_2^4 +12Y_X^2g_Y^2g_2^2 \nonumber \\
&\qquad
-12Y_X^2\lambda_X'g_Y^2 -105\lambda_X'g_2^2%\red{+0\times g_Y^4}
+{1\over2}\kappa'^2 -{7\over25}|\lambda_{{\Phi^\dagger}^2X^2}|^2
\bigg),\nonumber \\
%==========
&\frac{d\lambda_X''}{dt} = {1\over16\pi^2} \bigg(
+8\lambda_X'^2+24\lambda_X\lambda_X''+2\lambda_X'\lambda_X''+{1715\over2}\lambda_X''^2\nonumber \\
&\qquad
+6g_2^4 -105\lambda_X''g_2^2 -12Y_X^2\lambda_X'' g_Y^2
+{2\over25}|\lambda_{{\Phi^\dagger}^2X^2}|^2
%\red{+0\times g_Y^4+0\times g_Y^2g_2^2}
\bigg),\nonumber \\
%==========
&\frac{d\kappa}{dt} = {1\over16\pi^2} \bigg(
+3Y_X^2 g_Y^4 +{105\over4}g_2^4 \nonumber \\
&\qquad
+12\kappa\lambda +28\kappa\lambda_X 
+35\kappa\lambda_X' +{3535\over4}\kappa\lambda_X'' \nonumber \\
&\qquad
+6y_t^2\kappa -57\kappa g_2^2
-{3\over2}\kappa g_Y^2-6Y_X^2\kappa g_Y^2\nonumber \\
&\qquad
+4\kappa^2 +{35\over4}\kappa'^2
+{56\over5}|\lambda_{{\Phi^\dagger}^2X^2}|^2
\bigg),\nonumber \\
%==========
&\frac{d\kappa'}{dt} = {1\over16\pi^2} \bigg(
+12Y_Xg_Y^2g_2^2 \nonumber \\
&\qquad
+4\kappa'\lambda +4\kappa'\lambda_X +101\kappa'\lambda_X'
+{697\over4}\kappa'\lambda_X'' +6y_t^2\kappa' \nonumber \\
&\qquad
-{3\over2}\kappa'g_Y^2 -6Y_X^2\kappa'g_Y^2 -57\kappa'g_2^2 \nonumber \\
&\qquad
+8\kappa\kappa' +{64\over25}|\lambda_{{\Phi^\dagger}^2X^2}|^2
\bigg),\nonumber \\
%==========
&\frac{d\lambda_{{\Phi^\dagger}^2X^2}}{dt} = {\lambda_{{\Phi^\dagger}^2X^2}\over16\pi^2} \bigg(
+4\lambda +4\lambda_X -31\lambda_X' +{961\over4}\lambda_X'' \nonumber \\
&\qquad
+6y_t^2 -{3\over2}g_Y^2 -6Y_X^2g_Y^2 -57g_2^2 +4\kappa' +8\kappa
\bigg). 
%==========
\end{align}

%++++++++++++++++++++++++++++++++++++++++++++++++++++++++++++++++++++++
\item SM with a real septet scalar field
\begin{align}
%==========
&\frac{d\lambda}{dt} = {1\over16\pi^2} \bigg(
+24\lambda^2 -6y_t^4 +{9\over8}g_2^4 +{3\over8}g_Y^4 \nonumber \\
&\qquad
+12\lambda y_t^2 -3\lambda g_Y^2 -9\lambda g_2^2
+{3\over4}g_Y^2g_2^2 +{7\over2}\kappa^2
\bigg),\nonumber \\
%==========
&\frac{d\lambda_X}{dt} = {1\over16\pi^2} \big(
+30\lambda_X^2+2448\lambda_X\lambda_X''+51840\lambda_X''^2 \nonumber \\
&\qquad
-144\lambda_X g_2^2 +2\kappa^2
\big),\nonumber \\
%==========
&\frac{d\lambda_X''}{dt} = {1\over16\pi^2}\big(
+6g_2^4
+1530\lambda_X''^2+24\lambda_X\lambda_X''-144\lambda_X''g_2^2
\big),\nonumber \\
%==========
&\frac{d\kappa}{dt} = {1\over16\pi^2} \bigg(
+36g_2^4
+12\kappa\lambda +18\kappa\lambda_X +6y_t^2\kappa \nonumber \\
%+24\kappa\lambda_X'
&\qquad 
-{3\over2}\kappa g_Y^2 -{153\over2}\kappa g_2^2 +4\kappa^2
\bigg).
%==========
\end{align}

%++++++++++++++++++++++++++++++++++++++++++++++++++++++++++++++++++++++
\item SM with a complex septet scalar field
\begin{align}
%==========
&\frac{d\lambda}{dt} = {1\over16\pi^2} \bigg(
+ 24 \lambda^2 -6 y_t^4 +{3\over8} g_Y^4 + {9\over8} g_2^4 \nonumber \\
&\qquad
+ 12 y_t^2 \lambda - 9 \lambda g_2^2 -3 \lambda g_Y^2 + {3\over4} g_2^2 g_Y^2 
+7 \kappa^2 + 7 \kappa'^2 
\bigg),\nonumber \\
%==========
&\frac{d\lambda_X}{dt} = {1\over16\pi^2} \big(
 + 44 \lambda_{X}^2 + 96 \lambda_{X} \lambda_{X}' 
+3744 \lambda_{X} \lambda_{X}'' 
\nonumber \\
&\qquad
+ 77184 \lambda_{X}''^2 +9216 \lambda_{X} \lambda_{X}''' +31104 \lambda_{X}' \lambda_{X}''' \nonumber \\
&\qquad
+90432 \lambda_{X}'' \lambda_{X}''' + 2859264 \lambda_{X}'''^2 +6 Y_X^4 g_Y^4 \nonumber \\
&\qquad
- 12 Y_X^2 \lambda_{X} g_Y^2 -144 \lambda_{X} g_2^2 +2 \kappa^2
\big),\nonumber \\
%==========
&\frac{d\lambda_X'}{dt} = {1\over16\pi^2} \bigg(
+ 24 \lambda_{X} \lambda_{X}' + 204 \lambda_{X}'^2 +768 \lambda_{X}' \lambda_{X}'' \nonumber \\
&\qquad
+ 7008 \lambda_{X}''^2 +44616 \lambda_{X}' \lambda_{X}''' -73824 \lambda_{X}'' \lambda_{X}''' \nonumber \\
&\qquad
+ 2408676 \lambda_{X}'''^2 + 3 g_2^4 -12 Y_X^2 \lambda_{X}' g_Y^2 -144 \lambda_{X}' g_2^2 \nonumber \\
&\qquad
+ 12 Y_X^2 g_2^2 g_Y^2 +{1\over2}\kappa'^2 
\bigg),\nonumber \\
%==========
&\frac{d\lambda_X''}{dt} = {1\over16\pi^2} \big(
+8 \lambda_{X}'^2 + 24 \lambda_{X} \lambda_{X}'' +96 \lambda_{X}' \lambda_{X}'' \nonumber \\
&\qquad
+ 1588 \lambda_{X}''^2 +912 \lambda_{X}' \lambda_{X}''' +5312 \lambda_{X}'' \lambda_{X}''' 
+22696 \lambda_{X}'''^2 \nonumber \\
&\qquad
+ 6 g_2^4 -144 \lambda_{X}'' g_2^2 - 12 Y_X^2 \lambda_{X}'' g_Y^2 
\big),\nonumber \\
%==========
&\frac{d\lambda_X'''}{dt} = {1\over16\pi^2} \big(
+16 \lambda_{X}' \lambda_{X}'' - 80 \lambda_{X}''^2 +24 \lambda_{X} \lambda_{X}''' \nonumber \\
&\qquad
- 128 \lambda_{X}' \lambda_{X}''' +3056 \lambda_{X}'' \lambda_{X}''' - 12200 \lambda_{X}'''^2\nonumber \\
&\qquad
- 12 Y_X^2 \lambda_{X}''' g_Y^2 -144 \lambda_{X}''' g_2^2 \big),
\nonumber \\
%==========
&\frac{d\kappa}{dt} = {1\over16\pi^2} \bigg(
+ 3 Y_X^2 g_Y^4 + 36 g_2^4
+12 \kappa \lambda + 32 \kappa \lambda_{X} \nonumber \\
&\qquad
+48 \kappa \lambda_{X}' + 1872 \kappa \lambda_{X}'' +4608 \kappa \lambda_{X}''' 
+6 y_t^2 \kappa \nonumber \\
&\qquad
-{3\over2} \kappa g_Y^2 - 6 Y_X^2 \kappa g_Y^2 - {153\over2} \kappa g_2^2
+ 4 \kappa^2 + 12 \kappa'^2
\bigg),\nonumber \\
%==========
&\frac{d\kappa'}{dt} = {1\over16\pi^2} \bigg(
+ 12 Y_X g_2^2 g_Y^2 +4 \kappa' \lambda +4 \kappa' \lambda_{X} \nonumber \\
&\qquad
+ 156 \kappa' \lambda_{X}' +384 \kappa' \lambda_{X}'' + 13236 \kappa' \lambda_{X}''' 
+6 y_t^2 \kappa' \nonumber \\
&\qquad
- {3\over2} \kappa' g_Y^2 -6 Y_X^2 \kappa' g_Y^2 - {153\over2} \kappa' g_2^2 + 8 \kappa \kappa'
\bigg).
%==========
\end{align}
%++++++++++++++++++++++++++++++++++++++++++++++++++++++++++++++++++++++
\end{itemize}

\end{document}